**Gate Tunable Dissipation and "Superconductor-Insulator" Transition in Carbon Nanotube Josephson Transistors**

Gang Liu, Yong Zhang and Chun Ning Lau*

Department of Physics and Astronomy, University of California, Riverside, Riverside, CA 92521

* To whom correspondence should be addressed. lau@physics.ucr.edu

**Abstract**

Dissipation is ubiquitous in quantum systems, and its interplay with fluctuations is critical to maintaining quantum coherence. We experimentally investigate the dissipation dynamics in single-walled carbon nanotubes coupled to superconductors. The voltage-current characteristics display gate-tunable hysteresis, with sizes that perfectly correlate with the normal state resistance $R_N$, indicating the junction undergoes a periodic modulation between underdamped and overdamped regimes. Surprisingly, when a device's Fermi-level is tuned through a local conductance minimum, we observe a gate-controlled transition from superconducting-like to insulating-like states, with a "critical" $R_N$ value ~ 8-20 k$\Omega$.

The interplay between dissipation, disorder and fluctuations has been the focus of much theoretical work[1-8] in quantum bosonic systems. In particular, superconductor-insulator transition (SIT) observed in single Josephson junctions (JJ)[9], nanowires[10] [11], two-dimensional (2D) JJ arrays[12], and 2D thin films[13] have been attributed to changes in dissipation. However, dissipation in these systems are typically determined by device geometry and/or shunt resistors, and cannot be tuned *in situ*; thus direct tests of theoretical predictions are not straightforward because of inevitable variability in mesoscopic devices. Though dissipation and SIT via *in situ* control was reported in 2D JJ arrays fabricated on semiconductor heterostructures [12] and more recently in atomic Bose-Hubbard systems[14], such phenomena were not observed or examined in detail for a *single* JJ, the simplest Josephson system.

Here we report observation of gate tunable dissipation and evidence for a dissipation-driven SIT in individual single-walled carbon nanotube (SWNT) Josephson transistors[15-24]. As the devices are tuned by gate voltage from resonant to off-resonant transmission through the quantized energy levels, the critical current $I_c$ decreases significantly, while the voltage-current (*V-I*) characteristics change from non-hysteretic to strongly hysteretic. This indicates that the junction crosses over from overdamped to underdamped regimes, in agreement with estimates from the model of resistively and capacitively shunted junctions (RCSJ). Interestingly, as the Fermi level of each device is tuned through a local resistance maximum, we observe a "superconductor-insulator" transition, with a critical normal state resistance $R_N \sim 10$ k$\Omega$, similar to the threshold values observed in other superconducting systems. Our results demonstrate continuous modulation of damping in individual devices, and constitute the first observation of a gate-controlled SIT in a single JJ. Such SWNT JJs will enable *in situ* study of the interplay between dissipation and quantum coherence, and have implications for implementing quantum computation in superconducting systems.

The SWNTs are synthesized by chemical vapor deposition on highly doped Si substrates with a 300-nm thick $SiO_2$ dielectric layer [25], and contacted to electrodes (5nm of Pd and 80 nm of Al) (Fig. 1a). Similar results were observed in 5 different devices. For a typical device, the normal state linear-response conductance $G_N$ oscillates sinusoidally in gate voltage $V_g$ (Fig. 1b, upper panel). The data are taken at bias ~0.7mV to exclude the superconducting proximity effect. Its magnitude varies between 80 and 125 µS, approaching the theoretical limit of $G_Q=2e^2/h$ ≈160 µS for a perfectly coupled SWNT (here $e$ is the electron's charge and $h$ is the Planck's constant). The lower panel of Fig. 1b displays two-probe non-linear transport spectroscopy measurement of the same device at 260mK, plotting differential conductance ($dI/dV$) as a function of $V_g$ and applied bias. The striking checker-board pattern is a signature of Fabry-Perot interference of electron waves in the nanotube, whereas the conductance maxima (minima) correspond to the resonant (anti-resonant) transmission of charges via discrete energy levels, which arise from the finite length of the nanotube. The characteristic energy scale of the oscillation $eV_c$, indicated by the black dot in Fig. 1b, yields the level spacing $hv_F/2L$, where $v_F \sim 10^6$ m/s is the Fermi velocity, and $L$ is the distance traveled by charges between successive reflections. From Fig. 1b, we obtain $V_c$≈11.8mV, suggesting $L$≈175 nm. This is in excellent agreement with the measured source-drain spacing of 180 nm, indicating that scattering only occurs at the nanotube-electrode interfaces. These data indicate that our SWNT devices are clean, well-coupled to electrodes, and support ballistic and phase coherent charge transport.

For small $V<2\Delta$, the effect of the superconducting electrodes manifests as dramatically enhanced conductance ($\Delta$ is the superconducting energy gap of Al). The conductance peaks at zero bias reach a few mS $\gg G_Q$, indicating the presence of a supercurrent (Fig. 1b -c). At small but finite biases, we observe conductance peaks at sub-harmonic multiples of $2\Delta$ (Fig. 1c). These peaks, which arise from multiple Andreev reflections (MAR) of charges at the superconductor-SWNT interfaces[15, 27, 28], persist through the gate voltage range, yielding an estimated $2\Delta \sim$ 240 µeV.

We now focus on the *V-I* characteristics of the device in the current bias regime, in which two outer electrodes are used to inject currents, and we measure voltage across an inner-electrode pair. At small current bias $I <\sim$ nA, supercurrent is observed, characterized by a zero or very low voltage; when the bias exceeds a critical value $I_c$, the device switches from the supercurrent branch to a resistive branch, with a slope that approaches $1/G_N$ at high biases. Unlike a standard JJ, where the *V-I* curve changes only with temperature, a SWNT JJ displays dramatic variation in *V-I* characteristics with $V_g$, resulting from the tuning of the electrodes' Fermi levels with respect to the quantized energy levels in the nanotube. Three representative traces at different $V_g$ are shown in Fig. 2a, and *α, β* and *γ* corresponds to off-resonance, intermediate and on-resonance transmission, respectively.

For two superconductors coupled via discrete energy levels with 2 spin-degenerate channels, the critical current in the wide resonance regime, which is appropriate to our devices with extrinsic level widths $\Gamma >\sim$ meV, is[29]

$$I_c \approx \frac{2e\Delta}{\hbar} \tanh\left(\frac{\Delta}{2k_B T}\right)\left[1 - \sqrt{1 - \frac{G_N}{4e^2/h}}\right] \quad (1)$$

where the term in the square brackets takes into account of the asymmetric coupling between the SWNT and the two electrodes, and $k_B$ is Boltzman's constant. Thus Eq. (1) predicts that $I_c$ increases with increasing $G_N$. This is borne out in Fig. 2a: at resonant states with maximal $G_N$ (*γ* trace), the measured $I_c$ is largest, 8.3 nA; $I_c$ decreases with $G_N$, reaching a minimum of 1.1nA when the transmission is off-resonant (α trace) (see also Fig. 2b). Here we take the value of $I_c$ be that of the largest slope *dV/dI*.

The dependence of $I_c$ on gate voltage has been studied in previous experiments [15, 16, 18-24, 29]. A relatively unexplored aspect is the *shape* of the *V-I* curve, which contains a wealth of information on the junction dynamics. For instance, within the RCSJ model, the *V-I* characteristic of an underdamped junction in the strong coupling regime ($E_J >> k_B T$) is hysteretic, displaying sharp switching from supercurrent to resistive branches at $I_c$, and the reverse transition from resistive to supercurrent branches occurs at retrapping current $I_r \leq I_c$. Here $E_J = \frac{\hbar I_c}{2e}$ is the Josephson coupling energy. At finite temperatures, thermal fluctuations reduce $I_c$ but increase $I_r$, thereby reducing the hysteresis size, until $I_r = I_c$ at $E_J < \sim k_B T$. In comparison, in an overdamped junction the transition from supercurrent to resistive branches is *always* smooth and non-hysteretic. Thus, hysteresis observed in the *V-I* characteristics, or the lack thereof, is an important indication of the junction dynamics, as well as the relative magnitudes of Josephson coupling to thermal fluctuations.

Examining the three *V-I* curves in Fig. 2a, we observe that the α trace, which has the smallest $I_c$ and corresponds to off-resonance transmission, displays sharp switching in both current directions, with the largest hysteresis. For the β trace, hysteresis is small but observable. Both curves suggest relatively small dissipation in the junction. In contrast, the γ trace for on-resonance transmission has the largest $I_c$; it is also smooth and entirely non-hysteretic, resembling that of an overdamped junction. Thus, *when the transmission is tuned from on- to off-resonance, $I_c$ decreases while hysteresis increases.* Similar trend is observed over a large range in $V_g$: Fig. 2b displays the measured junction voltage (color) as functions of *I* (vertical axis) and

$V_g$ (horizontal axis) *for both directions of current-sweeping* – the up-sweep plot is set to be 50% transparent and superimposed on the down-sweep plot. The dark red and dark blue areas correspond to the resistive branches, and the white regions to the supercurrent branch. The hysteresis appear as the pale blue and pale red regions, bounded by the values of $I_c$ and $I_r$, with sizes that change with $V_g$ (note that in this representation, the pale blue/red regions do not correspond to those of low signals).

A quantitative analysis is presented in Fig. 3a. The top panel plots the gate dependence of $I_c$ (circles) and $I_r$ (filled circles), and the bottom panel plots the ratio $I_c/I_r$ that parameterizes the size of hysteresis (left axis), and $G_N$ (right axis) for ease of comparison. Clearly, the hysteresis size anti-correlates with $I_c$ and $G_N$: the ratio is ~ 1 when $G_N \sim 3e^2/h$, and increases to ~1.6 at $G_N \sim 2\ e^2/h$ (Fig. 3b). Naively, we expect that in an underdamped junction, the hysteresis size directly correlates with the ratio $E_J/kT$; hence large hysteresis should occur at on-resonant transmission, which is exactly opposite to our experimental observation.

Motivated by the observed anti-correlation of hysteresis and $G_N$, we consider the possibility of gate tunable damping in the SWNT JJ. Within the RCSJ model, dissipation can be parameterized by $1/Q$, where $Q$ is the junction's quality factor, given by[31]

$$Q = \omega_p R_j C_j, \qquad (2)$$

where $\omega_p = \sqrt{\dfrac{2eI_c}{\hbar C_j}}$ is the plasma frequency, $R_j$ and $C_j$ are the shunt resistance and capacitance of the junction, respectively. Thus $Q>1$ corresponds to underdamped regimes. In the absence of an external shunt resistor, $R_j \sim R_N$, and $C_j$ is estimated from geometric capacitance and gate conversion factor to be ~ 0.2 fF. For typical $R_N \sim 10$ kΩ, we obtain $Q$~1.3, indicating that the junction is close to being critically damped. This is consistent with our observation of transition from underdamped to overdamped regimes, as $G_N$ is modulated by the gate voltage. At $T=0$, $Q$ is

related to the ratio $I_c/I_r$ by[31]

$$I_c/I_r = \pi Q/4 \qquad (3)$$

In the presence of thermal fluctuations, the ratio is expected to decrease; nevertheless, in general $I_c/I_r$ can serve as a measure of dissipation in the junction at a fixed temperature. Combining Eq.s (1) to (3), we obtain $I_c/I_r \propto Q$, which is proportional to $R_N$ if the latter is sufficiently close to $h/4e^2$, and crosses over to $\sim\sqrt{R_N}$ with increasing $R_N$. Such trend is observed experimentally (Fig. 3b), though the crossover appears to occur at a lower value of $R_N$ than expected, which may be attributed to additional modification induced by thermal fluctuations. Thus, our results show that dissipation in the SWNT junction increases with $R_N$, which is continuously tunable by gate voltage.

The observation of gate tunable dissipation raises the possibility of observing superconductor-insulator transition (SIT) in SWNT Josephson junctions: for sufficiently small $G_N$, low dissipation may result in delocalization of the phase of the superconducting order parameter, *i.e.*, an insulating phase. This insulating phase is typically characterized by the presence of a large differential resistance at low temperature or small bias, which decreases to $R_N$ when $T$ or $I$ increases. Evidence for such a transition was observed in device 2 with a source-drain separation of 190 nm. For $V_g<9V$, it display behaviors very similar to those of device 1: regular Fabry-Perot patterns (Fig. 4a), MAR features which yield $2\Delta \sim 200$ μeV, and gate-tunable hysteretic *V-I* characteristics. However, at $V_g \sim 10.5V$, there exists a particularly resistive period with a minimum $G_N \sim 3\mu S$. Such a resistive state may arise from the presence of a small bandgap[31]. Within the gap, the *V-I* characteristics are dramatically different from those at other gate voltages: the zero bias resistance $R_0 \sim 3M\Omega \gg R_N$, and decreases to $R_N$ for $V>2\Delta \sim 210$ μeV (Fig. 4b, red curve). Such negative *dR/dI* resembles that of the insulating phase, in stark contrast with the typical *V-I* characteristics of a superconducting junction shown in Fig. 2, in which a small $R_0$ increases to $R_N$ at high current biases. This insulator-like behavior within the gap is further verified by the temperature dependence of the *V-I* characteristics[34]. Fig. 4b plots the *V-I* curves at $V_g$ =10.5V at 260mK, 530mK and 700mK, respectively, showing deceasing $R_0$ with increasing temperature; the larger bias *V-I* curves are shown in the upper panel of Fig. 4c. Note that $R_N$, *i.e.*, the slope of the *V-I* curve at high biases, stays constant for all three temperatures, indicating that the junction remains metallic, and the negative $R_0$-*T* dependence arises collectively from the S/SWNT/S junction. For comparison, the *V-I* curves outside the gap at different temperatures are shown in the lower panel of Fig. 4c, displaying $R_0$ that increases with higher *T*.

Further evidence for a gate-tunable SIT is the temperature dependence of $R_0$ of the junction at a number of $V_g$ with different $R_N$, as shown in Fig. 4d. At *T*~ 1K, the *V-I* characteristics are linear, and $R_0 = R_N$. As temperature is decreased, depending on the value of $R_N$, a clear divergence of behavior is evident: at gate voltages with $R_N$~10kΩ, $R_0$ decreases with temperature and approaches 0 at 260mK; when $R_N$>20kΩ, $R_0$ increases with temperature; those gate voltages with intermediate $R_N$ have almost constant $R_0(T)$. This dichotomy of behavior is clearly visible despite the scatter in the data points, which arises from shifts in the effective $V_g$ due to random fluctuation in the charge environment. Thus the junction may be superconductor-like or insulator-like, depending on a "critical" value of $R_N$ ~8-20 kΩ. Such behavior is reminiscent of the dissipation-driven SIT observed in 2D superconducting thin films and JJ arrays, where the transition occurs when the sheet resistance is equal to $h/4e^2$~6.5kΩ[9, 12, 33]. In the context of gate tunable dissipation described in the first part of the letter, the SIT is likely driven by modulations in dissipation, though the detailed dynamics and dissipation mechanism warrant further experimental and theoretical investigation.

In conclusion, we observed gate tunable dissipation in SWNTs JJs, and the first gate-modulated superconductor-insulator transition a single JJ. As SWNTs continue to offer a unique platform for exploring the interplay between dissipation and quantum coherence, further experimental investigation will be necessary to pinpoint the critical value of $R_N$ and the dissipation mechanism.

We thank Leonid Pryadko, Gil Refael and Marc Bockrath for stimulating discussions, and Hsin-Ying Chiu for helpful suggestions for fabrication. This research is supported in part by NSF CAREER DMR/0748910 and ONR/DMEA Award H94003-07-2-0703.

34. We observe that the values of $\Delta(T)$ obtained from the $V$-$I$ curves exhibits the BCS dependence on temperature.

Fig. 1.(a). Scanning electron micrograph of a nanotube (gold) connected to 8 electrodes (white). Scale bar: 400 nm. (b) Upper panel: Normal state conductance of a SWNT device as a function of $V_g$ at 260mK at $V\sim0.7$mV. Lower panel: Differential conductance *vs.* bias $V$ and gate voltages $V_g$. (c). High resolution plot of *dI/dV* vs. $V$ and $V_g$ at small biases. Notice the logarithmic color scale.

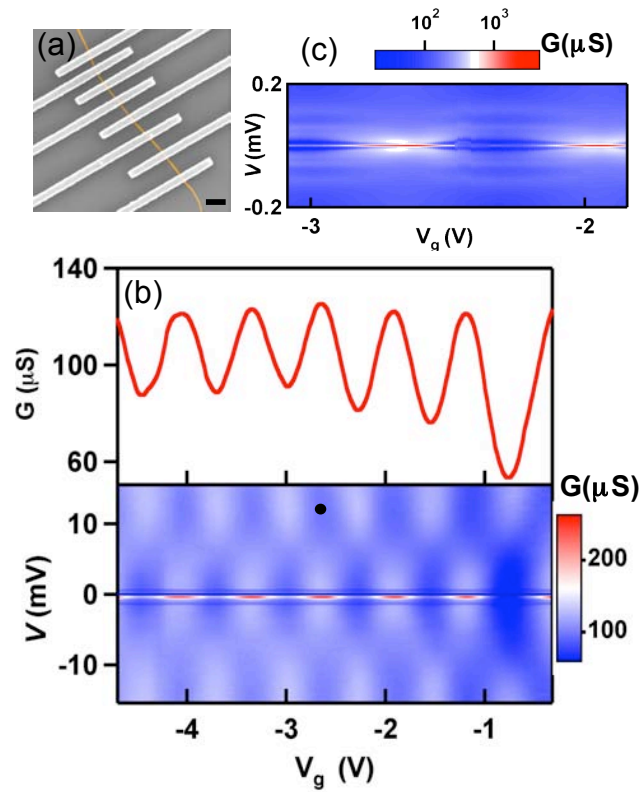

Fig. 2. (a). *V-I* characteristics of a S/SWNT/S device at 260mK at 3 different $V_g$. Inset: γ curve at large current bias. (b) Measured voltage across the device *vs.* *I* and $V_g$, indicating transition from supercurrent (white) to resistive (blue/red) states. α, β and γ indicate the gate voltages at which line traces in (a) are taken. The figure consists of data taken in both up-sweep and down-sweep directions, which are set to be 50% transparent and superimposed together to accentuate the gate-dependent hysteresis (pale blue/red).

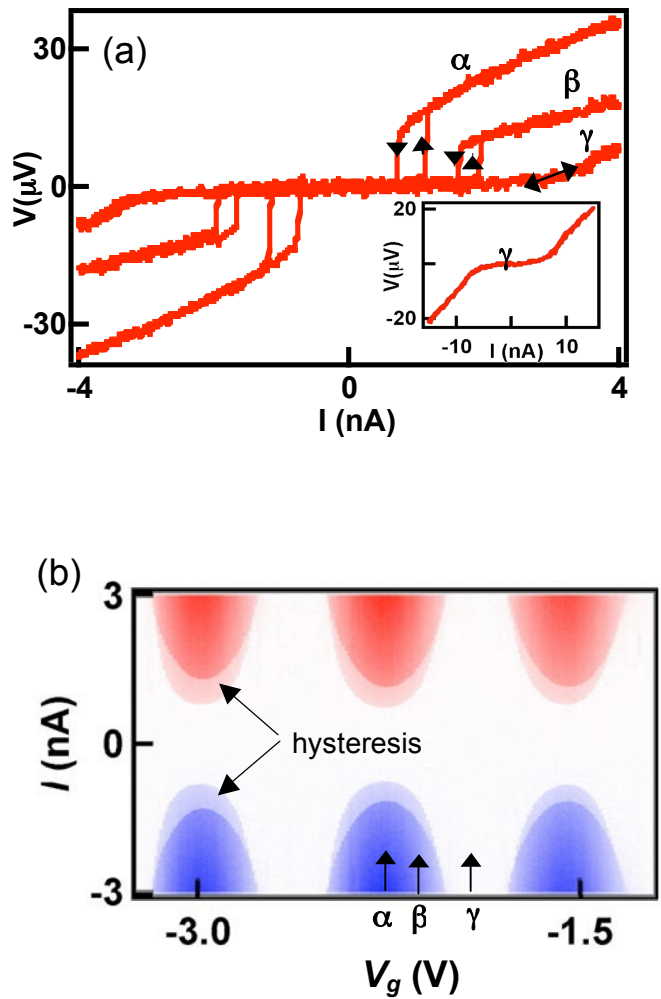

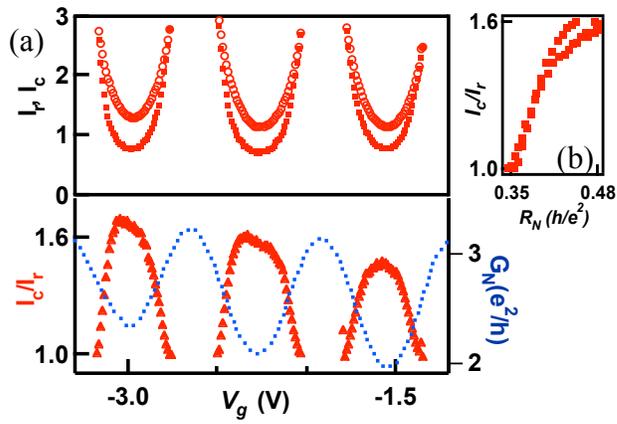

Fig. 3. (a). Upper panel: $I_c$ (open circles) and $I_r$ (closed circles) vs. $V_g$. Lower Panel: $I_c/I_r$ (triangles, left axis) and $G_N$ (dotted line, right axis) vs $V_g$. (b). $I_c/I_r$ vs. $R_N$ for $V_g = -2$ to $-2.5$V.

Fig. 4. (a). $G_N$ vs $V_g$ for a device with a conductance minimum at $V_g \sim 10.5$V. (b). V-I characteristics at $V_g=10.5$V at $T=0.7$K (red), 0.5K (green) and 0.26K (blue). (c). Upper Panel: same as (b) with larger bias range. Lower panel: V-I curves at $V_g=-0.51$V at $T=0.9$K (red), 0.53K (green) and 0.26K (blue). (d). $R_0$ vs $T$ at different $V_g$ (from top to bottom, $V_g$ = 10.5, 10.72, 10.83, 11, -0.81, -0.71, -0.61, -0.58 and -0.55 V).

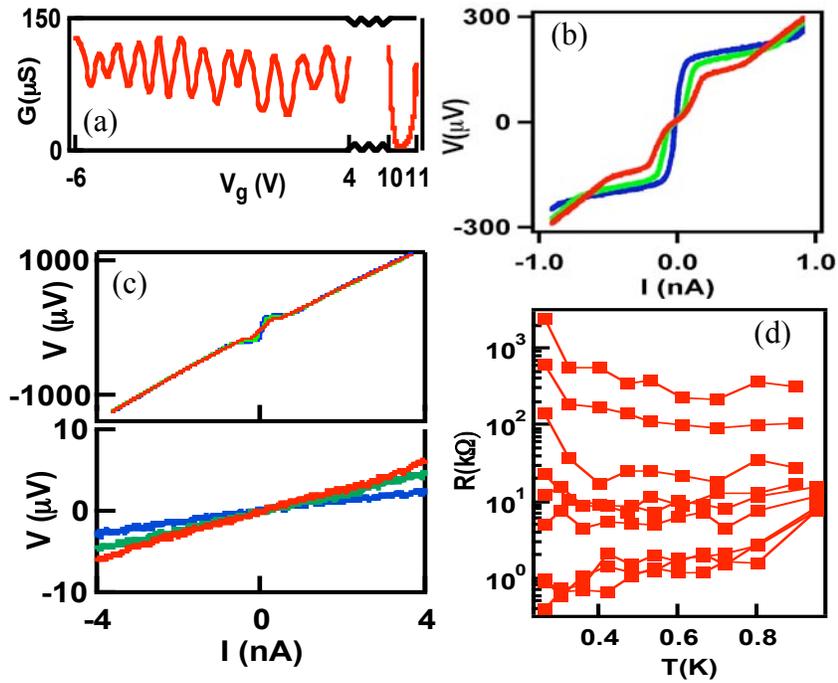